# Coexistence of large anomalous Hall effect and topological magnetic skyrmions in a Weyl nodal ring ferromagnet $Mn_5Ge_3$

Hang Li[1,5], Feng Zhou[1,5], Bei Ding[2,5], Jie Chen[1], Linxuan Song[2], Wenyun Yang[4], Yong-Chang Lau[2], Jinbo Yang[4], Yue Li[1], Yong Jiang[1], Wenhong Wang[1,2]*

[1]Institute of Quantum Materials and Devices, School of Electronics and Information Engineering, Tiangong University, Tianjin 300387, China
[2]Institute of Physics, Chinese Academy of Sciences, Beijing 100190, China
[3]University of Chinese Academy of Sciences, Beijing 100049, China
[4]State Key Laboratory for Artificial Microstructure & Mesoscopic Physics, School of Physics, Peking University, Beijing 100871, P. R. China
[5] These authors contributed equally: Hang Li, Feng Zhou, and Bei Ding.
*E-mail: wenhongwang@tiangong.edu.cn

**Topological magnetic materials are expected to show multiple transport responses because of their unusual bulk electronic topology in momentum space and topological spin texture in real space. However, such multiple topological properties-hosting materials are rare in nature. In this work, we reveal the coexistence of a large tunable anomalous Hall effect and topological magnetic skyrmions in a Weyl nodal ring ferromagnet $Mn_5Ge_3$, by using electrical transport and Lorentz transmission electronic microscope (TEM) measurements. It was found that the intrinsic anomalous Hall conductivity (AHC) can reach up to 979.7 S/cm with current along [120] and magnetic field along [001] of the $Mn_5Ge_3$ single crystals. Our theoretical calculations reveal that the large AHC is closely related with two Weyl nodal rings in band structure near the Fermi level and is strongly modified by the content of Ge. Moreover, our Lorentz-TEM images and micromagnetic simulation results, together with the sizable topological Hall effect clearly point to the robust formation of magnetic skyrmions over a wide temperature-magnetic field region. These results prove $Mn_5Ge_3$ as a rare magnetic topological nodal-line semimetal with great significance to explore novel multiple topological phenomena, which facilitate the development of spintronics.**

Magnetic topological materials have recently attracted substantial research attention due to the interplay between the nontrivial band structures and spin degrees of freedom that offer the potential for revolutionary spintronic applications[1, 2]. In a magnetic topological system, many transport responses, in particular, anomalous Hall conductivity (AHC) can be significantly enhanced due to the influence of the Berry curvature that is associated with nontrivial bands in momentum space[3, 4]. This has recently been recognized in various topological magnetic systems, including the chiral antiferromagnets $Mn_3Sn$[5] and $Mn_3Ge$[6], kagome lattice ferromagnets $Fe_3Sn_2$[7] and $Co_3Sn_2S_2$[8], nodal-line ferromagnets $Fe_3GeTe_2$[9], $Co_2MnZ$ (Z = Ga, Al)[10, 11, 12, 13], $MnAlGe$[14] and recently in two TiNiSi-type Transition metal pnictides ZrMnP and HfMnP[15]. These materials have shown exciting transport responses, including extremely large anomalous Hall effect, giant magnetoresistance (MR), and chiral anomaly[2]. Meanwhile, besides the nontrivial band topology in momentum space, the spin structure topology in real space can also cause numerous exotic transport properties and spectroscopic behaviors. For example, the presence of topologically protected vortex-like spin textures, i.e., magnetic skyrmions, has been revealed experimentally in various bulk magnetic materials, ranging from non-centrosymmetric chiral magnets, such as $MnSi$[16] and $FeGe$[17], to centrosymmetric magnets including $MnNiGa$[18] and $Fe_3Sn_2$[19]. More importantly, these topological properties observed either in momentum space or in real space may coexist in one material simultaneously. For instance, the commonality of the nontrivial band structure in reciprocal space and spin textures in real space is that they are accompanied by large Berry curvatures and can lead to new topological phases. These materials facilitate the observation of large anomalous electrical and thermal transport features. Therefore, materials with multiple topological magnetic properties are crucial for the advancement of topological research and practical applications. To date, however, multiple topological magnetic materials, i.e., the coexistence of large intrinsic AHE and topological magnetic skyrmions, have been experimentally realized only in abovementioned two ferromagnets, i.e., kagome $Fe_3Sn_2$[19] and layered $Fe_3GeTe_2$[20]. Therefore, to search for new magnetic topological materials showing both large intrinsic AHE and topological magnetic skyrmions will achieve advantages in broadening the material platform for future spintronics as well as for the quest toward novel topological phenomena.

In this work, we unambiguously reveal the coexistence of large anomalous Hall effect and magnetic skyrmions in $Mn_5Ge_3$ single crystals through detailed transport measurements and Lorentz transmission electron microscopy (TEM) combined with *ab*-initio calculations. This compound has attracted much attention due to its significant magnetocaloric effect[21] and anomalous Hall effect[22], as well as its recent prediction of topological Dirac magnons[23]. There also are some studies focus on the element doping to improve its Curie temperature ($T_C$)[24, 25, 26, 27, 28], and the large spin polarization[29] and compatibility with current semiconductor processes also makes the material a candidate for the preparation of novel germanium-based electrical devices[30, 31, 32]. Here, we theoretically predict that $Mn_5Ge_3$ is a Weyl nodal ring (NR) ferromagnet. From the comparison between our experimental results and theoretical calculations, we confirm that two Weyl NRs induce a strong Berry curvature near the Fermi level, which is responsible for the observed large intrinsic anomalous Hall conductivity (AHC) of ~ 979.7 S/cm. In addition, our Lorentz TEM experiments show that the magnetic skyrmions appear in *ab* plane when an appropriate magnetic field is applied along the [001] direction, and a skyrmion-induced topological Hall resistivity as large as ~ 972 nΩ cm is observed at 200 K. Our results provide a rare multiple topological magnetic material for simultaneously studying both the topological properties of inverse space and real space, which can be used in next generation information storage and logic computing devices

## Results and Discussion

**Crystal/magnetic/electronic band structures**

As shown in Fig. 1a, $Mn_5Ge_3$ crystallizes a hexagonal $D8_8$-type crystal structure with space group $P6_3/mcm$ (no. 193)[33, 34], therefore indicating the presence of several mirror symmetries in the system. This is crucial for lifting spin degeneracy and necessary for forming topological nodal rings. The Mn atoms locate in two different positions Mn1 (0.2335, 0, 0.25) and Mn2 (1/3, 2/3, 0), as shown purple and orange balls. Fig. 1b is the neutron powder diffraction data at 5K, the Rietveld refinement results shown that the magnetic moments of $Mn_5Ge_3$ are colinear and aligned along the [001] (*c*-axis) direction (as shown in Fig. 1c), and the Mn atoms are the main magnetic contributors, with the magnetic moments of about 3.32 and 2.25 μ$_B$ for Mn1 and Mn2, which is consistent with the former results[35]. The Ge atoms (green balls) are in the same layer as Mn1 atoms and form triangular networks (top view in Fig. 1a). The Mn$_5$Ge$_3$ lattice can also be viewed as a Mn2 layer intercalating two adjacent Mn1+Ge layers (side view in Fig. 1a). This particular stacking is reflection symmetric with respect to the Mn1+Ge layers. In other words, the Mn1+Ge atomic planes are a mirror plane of the crystal lattice under the mirror operation, which provides a protection for the topological nodal rings, as we will discuss later on.

We first performed density functional theory (DFT) calculations on the band structures and anomalous Hall conductivity (AHC) of Mn$_5$Ge$_3$ (see the "Methods" section). Fig. 1d shows the band structure of Mn$_5$Ge$_3$ without spin-orbit coupling (SOC). There are eight band crossing points (BCPs) named P1, P1', P2, P2', P3, P3', P4, and P4' in the Γ-M-K-Γ paths. The BCPs P2 and P2' are protected by the M$_z$ symmetry, while the other BCPs cross by opposite spin bands and are protected by the U(1) symmetry of spin rotation. Therefore, they are not isolated nodes but a closed nodal ring (NR) in the k$_z$=0 plane. These four NRs in k$_z$=0 plane, which are denoted by NR1, NR2, NR3, and NR4 are induced by band crossing can be found in Supplementary Fig. S1b. During considering the SOC effect, as we shown in Fig. 1e, the magnetic moments along [001] (*c*-axis), both of the NR1 and NR4 opened a small gap due to the band hybridization. However, the NR2 and NR3 are survived by being protected from the symmetry of M$_z$. To describe the position and morphology of NR2 and NR3, we calculated the 3D/2D-band structure of the k$_z$=0 plane. Fig. 1f and g are the 3D and 2D schematics of the two energy bands that give rise to nodal rings, the green and black closed curves represent the two nodal rings NR2 and NR3, respectively. Fig. 1h is the contour plots of Berry curvature. It is clearly shown that the positions of two nodal rings correspond to high values of Berry curvature, which may result in a large intrinsic AHC. As shown in Fig. 1i, the calculated anomalous Hall conductivity ($\sigma_H^A$) is about 841 S/cm, in excellent agreement with the previous report of 860 S/cm in thin film of Mn$_5$Ge$_3$[22]. We note that, when magnetic moments are along the *ab*-plane, the $M_z$ symmetry is broken, and as a result, the NR2 and NR3 will be gapped (see Supplementary Fig. S2), suggesting the existence of the large anisotropic $\sigma_H^A$. Motivated by these considerations, we investigate bulk $Mn_5Ge_3$ single crystals by magneto-transport measurements as we will show later.

**Characterization of $Mn_5Ge_3$ single crystals**

The $Mn_5Ge_3$ single crystals used in this study were grown by using Ge self-flux method (see the "Methods" section). The samples were of high structural quality, which was confirmed by our X-ray diffraction and scanning transmission electron microscopy (STEM) measurements. An optical image of typical single crystals are hexagonal prisms as shown in the inset of **Fig. 2**a. The upper X-ray diffraction (XRD) pattern shown in Fig. 2a is the data from the rectangular side of the hexagonal prism-shaped single crystal. All the peaks are consistent with the (*h00*) series peaks of $Mn_5Ge_3$. The XRD pattern of the hexagonal section is shown below, which also correspond to the (*00l*) peaks of $Mn_5Ge_3$. The single-crystal X-ray diffraction patten (see Supplementary Fig. S3) delivers the same crystal structure as previous polycrystalline $Mn_5Ge_3$ powder neutron diffraction (PND) data[35].

To further confirm the chemical composition of our samples, we performed the energy dispersive analysis spectrometer (EDS) of X-rays and found that the as-grown single crystal is the Ge-rich non-stoichiometric $Mn_5Ge_3$ (see Supplementary Fig. S4). This is further corroborated by scanning transmission electron microscopy (STEM) images. As shown in Fig. 2b, c and d, the high-angle annular dark-field STEM (HAADF-STEM) image taken along the *ab* plane can be perfectly overlaid with the structural model. By comparing with the STEM and standard simulated images of Ge-rich $Mn_5Ge_3$ single crystals in *ab* plane, as shown in the Supplementary Fig. S5, we can see that the centers of hexagonal networks, (0,0,0) position in the $Mn_5Ge_3$ lattice, are occupied by the excess Ge atoms. The HAADF-STEM and simulated image taken along the *ac* plane, as shown in Supplementary Fig. S6, further suggest that these additional Ge atomic signals are random. Combined with the EDS results and the Rietveld refinements of single crystal powder XRD data (in Supplementary

Fig. S7 and S8), we can conclude that the as-grown single crystal is the Ge-rich Mn$_5$Ge$_3$. Next, we still use 'Mn$_5$Ge$_3$' to refer to our crystals for clarity.

The magnetic properties of Mn$_5$Ge$_3$ single crystals were characterized through magnetization M measurements (see Supplementary Fig. S9). Fig. 2e shows the typical M(T) curves measured in the absence of a magnetic field [zero-field–cooling (ZFC)] and cooled under a constant magnetic field of H=100 Oe oriented along the [001] direction [field-cooling (FC)]. As shown in the enlarged area of the black dashed rectangle, a slight decrease in magnetization curves M(T) at T~140 K may result from a random spin reorientation, which also appeared in other M(T) curves along other axes (see Supplementary Fig. S9). The high-temperature behavior is similar in both the ZFC and FC data and indicates the Curie temperature (T$_C$) at T~285K. Fig. 2f shows the isothermal magnetization M(H) curves at 10K measured with the magnetic field applied along [001], [120] and [100] directions, respectively. Magnetic induction density $B = \mu_0[H + (1 - N_d)M]$, where $\mu_0$ is vacuum permeability and $N_d$ is shape demagnetization factor. The sample used in this paper are rectangular whose $N_d$ can be calculated[36]. The values of $N_d$ along [001], [120] and [100] directions, are 0.178, 0.603 and 0.219, respectively. The temperature dependence of magnetic anisotropic constant (Ku) is displayed in the inset of Fig. 2f, which also shown a slight change near 140K. These magnetization data reveal that Mn$_5$Ge$_3$ single crystals are ferromagnetic properties and the easy magnetization is along [001] (c-axis), which is consistent with previous report on polycrystalline Mn$_5$Ge$_3$ sample confirmed PND experiments[37]. In addition, as shown in Fig. 2g, the temperature dependence of the zero-field longitudinal resistivity of Mn$_5$Ge$_3$ single crystals show clearly anisotropic behavior between the c-axis and ab-plane.

**Large anomalous Hall effect**

In general, for a magnetic conductor, the magnetic field dependence of Hall resistivity $\rho_H$ can be expressed as $\rho_H = \rho_H^N + \rho_H^A = R_0H + 4\pi R_s M$, where R$_0$, H and R$_s$ is normal Hall coefficient, magnetic field and anomalous Hall coefficient, respectively. In **Fig. 3**a, we show the magnetic field H dependence of $\rho_H$ curves measured at 10 K for Mn$_5$Ge$_3$ single crystals with magnetic field $H$ and current $I$ along different axes. More detailed electrical transport data are shown in Supplementary Fig. S10-13. Clearly, the measured $\rho_H$ curves show a similar magnetic field dependence as the M(H) curves presented in Fig. 2d, indicating that the anomalous Hall resistivity $\rho_H^A$ contribution (proportional to *M*) is dominant while the normal Hall resistivity $\rho_H^N$ contribution is very small. These curves all show a spontaneous value, by linear fitting of the high-field data (as shown by the purple dashed line in Fig. 3a), it is possible to obtain the $\rho_H^A$ values at zero filed. As shown in Fig. 3b, the extracted $\rho_H^A$ show strong temperature dependence for all directions and reach maximum values near 200 K. In addition, as shown in Fig. S10 and S11, the signs of Hall coefficient R$_0$ change at T~150 K during H along [001], which indicates the change in majority carrier types. This temperature is consistent with the slight spin reorientation in M(T) curves (see Supplementary Fig. S9). The phenomenon also observed in Mn$_5$Ge$_3$ films[22]. Correspondingly, the curves of Hall conductivity, $\sigma_H^A = \rho_H^A/(\rho^2 + \rho_H^2)$, for $H$ along the [001] (c-axis) and [120] (*ab*-plane) shows in Fig. 3c. As we expect from the DFT calculations, $\sigma_H^A$ shows a clear anisotropic behavior. We found that, the $\sigma_H^A$ can be as large as 1050 S/cm below 100 K with *I* and *H* along [120] and [001], respectively. More importantly, the different $\sigma_H^A$ between *I* along [120] or [100] and *H* along [001] indicates that there is an anisotropic electrical property in *ab* plane. To further study the contribution from intrinsic mechanism, as shown in Fig. 3(d), we extracted the anomalous Hall resistivity $\rho_H^A$ and the zero-field resistivity $\rho$, and plotted the data for $\rho_H^A$ V.S. $\rho^2$. The slops $\gamma$ of the linear fitting of α+γρ² represent the intrinsic contribution[38]. The intrinsic AHC is up to 979.7 S/cm during *I* along [120] and *H* along [001].

To discuss what physical parameters mainly contribute to the large anisotropic $\sigma_H^A$ in terms of the Berry curvature, we next applied an atomic description of composition changes using a density functional theory approach to gain a fundamental picture of such a doping-induced AHC modulation. We have calculated the band structures of Mn$_5$Ge$_{3+x}$ (x=0.25, 0.33, 0.5, 0.75 and 1.0) with the excess Ge atoms occupied randomly in the 2b Wyckoff sites (0,0,0) (see **Fig. 4**a-c and

Supplementary Fig. S14). As a result, we found that, with increasing of Ge composition in $Mn_5Ge_{3+x}$, the excess Ge atoms will provide extra electrons to occupy the valence band and thus moves $E_F$ to higher energies, approaching to the two NRs. However, as the amount of Ge increases (x > 0.5), the band structures are obvious changes and the Weyl NRs also dropped down below $E_F$ or disappeared.

One of the critical factors responsible for $E_F$ approaching to the Weyl NRs and thus endowing the Berry curvature with a large AHC is the excess Ge content x. It can be further seen that, as shown in Fig. 4d and 4e, the calculated AHC value of $Mn_5Ge_{3+x}$ near $E_F$ first increases and then decrease gradually with increasing of Ge content x. As a result, the main factor affecting the calculated AHC is the $E_F$ movement, leading to a sizable enhancement at x<0.5 and then decreases with a further increase in the Ge content x. In comparison, our supercell approach predicted a maximum value of AHC at x=0.5, indicating that the modulation of the electronic structure via Fermi-level tunning is indeed important to enhance the Berry curvature. The corresponding calculated value of AHC at x=0.33 is comparable with that of the intrinsic AHC (the red ball in Fig. 4e) measured from our Ge-rich $Mn_5Ge_3$ single crystals.

On the other hand, our supercell approach suggests that the $E_F$ shifting effect, which results in the $E_F$ approaching to the Weyl NRs tends to enhance the intrinsic contribution to the AHC of $Mn_5Ge_{3+x}$, as illustrated in Fig. 4f. Similar supercell approaches were applied to electron-doped magnetic Weyl semimetals $Co_{3-x}Ni_xSn_2S_2$ alloys, where the $E_F$-shifting and band broadening effect from local disorder are responsible for the enhancement of AHC[39]. In fact, there are several possible routes by which $E_F$ may be engineered, so that the topological transport properties can be tuned for spintronic devices. Firstly, carbon implantation has been explored in the literature to enhance $T_C$, but the effect on the Weyl NRs and on the Berry curvature is unknown. Secondly, chemical substitution of Ge by Si carbon implantation is demonstrated to connect to the multifunctional antiferromagnetic $Mn_5Si_3$. On the other hand, substituting Ge by Si reduces $T_C$, but once again the impact on the Weyl NRs and on the Berry curvature remains unexplored. Lastly, $Mn_5Ge_3$ can also be grown in thin film form with different Ge contents. The bulk band structures and the Berry curvature can be further modified by epitaxial strain.

**Topological spin textures**

Next, we utilized Lorentz transmission electron microscopy (L-TEM) to observe the magnetic domain structures in $Mn_5Ge_3$ single crystals. The lamellar samples used in L-TEM experiments were prepared by the mechanical polishing procedure and the sample surface was along the *ab* plane of the crystal. Fig.s 5a show the evolution of spin textures under vertically applied magnetic fields at 200 K after ZFC (more detailed L-TEM data are shown in Supplementary Fig. S15). Clearly, the magnetic stripe domains gradually shrink with field increasing, and when H= 270 mT, the ends of some strip domains begin to detach and transform into particle-like spin textures, i.e., magnetic skyrmion bubbles. When H=400 mT, only magnetic skyrmions are present in the images. As the field continues to increase, the bubbles will gradually shrink and eventually disappear, as is the case when H=600 mT. In addition, we noted that there are two different types of skyrmion bubbles in the image, marked with blue and green dotted rectangular boxes and red numbers 1 and 2, respectively. According to the previous reports in MnNiGa[40], the two skyrmions are type 1 and 2 bubbles with topological number n=+1 and -1, respectively. **Fig. 5**b exhibit the magnified images and chiral schematic of two magnetic skyrmions, which are captured from the corresponding dotted rectangular areas in Fig. 5a. Fig. 5c shown the micro-magnetic simulations results at 200 K (more detailed simulation data are shown in Supplementary Fig. S16), the patterns outlined by blue and red, showing concentric circular rings with alternating bright and dark contrasts, are typical Bloch-type skyrmions with opposite helicities. The opposite helicities coexist, which is consistent with the case of dipole skyrmions, meaning that the mechanism is also the competition among the perpendicular magnetic anisotropy, exchange, and dipolar interactions[18, 19]. From the whole field-dependent spin texture evolution at different temperatures in Supplementary Fig. S15, we found that the stable magnetic skyrmions can exist up to 260 K.

In addition, field-free magnetic skyrmions can be obtained via the FC process. Fig. 5d shows the under-focus L-TEM domain images in the zero-field at different temperatures after FC with H=500 Oe. Similar to the previous studies, the detailed FC procedure is schematically illustrated in Supplementary Fig. S17. Compared to the ZFC process, there is a significant increase in density of skyrmions. The stable temperature of these topological domains is increased by 10 K to 270 K. The diameter of skyrmions increases slightly with increasing temperature, which is slightly larger than the ZFC process. In the vicinity of $T_C$, as shown in the red arrows in Fig. 5d, the chirality of some skyrmions undergoes a spontaneous reversal, which is induced by the thermal disturbance[41].

**Topological Hall effect.**

The creation of magnetic skyrmions provides a basis for the emergence of topological Hall effect (THE) in $Mn_5Ge_3$ single crystals. We further analyzed the Hall resistivity between 5 and 300 K by considering the total Hall resistivity $\rho_H$ as the sum of three terms, $\rho_H = \rho_H^N + \rho_H^A + \rho_H^T$, where $\rho_H^T$ is the THE. Fig. 5e shows the topological Hall resistivity $\rho_H^T(H)$ curves at various temperature after subtracting the $\rho_H^N$ and $\rho_H^A$ parts. The process of extracting THE is shown in the inset of Fig. 5f. Remarkably, we obtained a considerable THE contribution which was indicated by the peaks on the $\rho_H^T(H)$ curve (blue). The non-zero THE from 5 to 300 K suggests that the topological domains can be stable over such a wide temperature range. Therefore, we propose that magnetic skyrmions may stable down to the low-T region. Unfortunately, the lowest temperature we can reach in L-TEM experiments is around 100K, which excludes the possibility of in-situ observation. All the $\rho_H^T(H)$ curves at 5–300 K were summarized as the color map in Fig. 5f. The dotted lines and hollow circles represent critical fields from THE and L-TEM, respectively, which is very similar to the phase diagram commonly observed in bulk magnets[42, 43]. Therefore, we conclude that the THE dominant in these regions may derived from magnetic skyrmions.

Our systematic transport measurements and Lorentz TEM experiments, supported by ab initio calculation, clearly indicate the coexistence of a large AHC and topological magnetic skyrmions in a ferromagnetic Weyl NRs candidate, $Mn_5Ge_3$. By measurement anomalous Hall resistivities, we got a large intrinsic anomalous Hall conductivity (AHC) about 979.7 S/cm with current along [120] and magnetic field H along [001] in Ge-rich $Mn_5Ge_3$ single crystals. The *ab* initio calculation indicated that the large AHC is closely related with two Weyl nodal rings in band structure near the Fermi level. Except for the topological structure in inverse space, the L-TEM experiments shown that the skyrmions will appeared in *ab* plane when an appropriate magnetic field is applied along the [001] and we also measured the topological Hall effect originated from these topological magnetic domains. Our results provide a rare magnetic topological material for simultaneously studying both the topological properties of inverse space and real space, this may enable the further exploration for new topological quantum states and promising used in future spintronics devices.

## Methods

**Sample preparation**

Single crystals of $Mn_5Ge_3$ were prepared by Ge self-flux method with a molar ratio of Mn:Ge= 11:9. Mn (99.9% purity) and Ge (99.999% purity) granules were put in an alumina crucible and then sealed in into vacuum quartz tube ($<5\times10^{-4}$ Pa). The quartz tube was placed in a furnace and kept at 1150°C for 1day to ensure that Mn is fully dissolved in Ge solution. The furnace is then slowly reduced to 850 K at a rate of 2 K/h and held at 850 K for 10 hours. Finally, the quartz tube was inserted upside down into the centrifuge to separate the single crystals and the excess Ge flux. $Mn_5Ge_3$ Polycrystals were prepared by arc-melt. The stoichiometric Mn (99.9% purity) and Ge (99.999% purity) granules were melted in vacuum arc furnace. Then the melt ingots were sealed in vacuum quartz tube ($<5\times10^{-4}$ Pa) and annealed at 850°C for 3 days for homogenization. Finally, the quartz tube was quickly drop into the ice-water mixture.

**Structural and composition characterizations**

The single crystallinity and orientation of the as-grown single crystal were determined using single-crystal X-ray diffraction (RAPID, Rigaku) with Cu Kα radiation at room temperature. The lattice parameters are obtained by Rietveld refinement. All the samples were shown to be single phase, with lattice parameters approximating previous polycrystals work. According to the energy dispersive X-ray analysis with a scanning electron microscope, the composition of single crystal can be determined to be Ge-rich $Mn_5Ge_3$. The powder neutron diffraction (PND) data were collected from Peking University High Intensity Powder Diffractometer (PKU-HIPD) at China Institute of Atomic Energy (CIAE). The polycrystal powder of $Mn_5Ge_3$ was loaded in a cylindrical vanadium sample holders and mounted in a sample changer for measurement. A Si(115) monochromator was used to produce a monochromatic neutron beam of wavelength 1.478 Å.

**Magnetic and transport characterizations**

Magnetic and electrical measurement used the Physical Properties Measurement System (PPMS) of Quantum Design Corporation. To calculated the demagnetization factors, magnetic measurement samples are rectangular bars prepared by cutting and polishing. The electrical measurements used standard six probes method to measure the resistance and Hall resistance at the same time. In order to eliminate the influence of probes misalignment, we measured the values at positive and negative filed and processed the resistivity and Hall resistivity by the formulas of $\rho(H) = [\rho(+H) + \rho(-H)]/2$ and $\rho_H(H) = [\rho_H(+H) - \rho_H(-H)]/2$, respectively. The Hall conductivity calculated by the function of $\sigma_H^A = \rho_H/(\rho^2 + \rho_H^2)$.

**Ab initio computation details**

Electronic structure calculations are performed in the Vienna ab initio Simulation Package (VASP) based on Density functional theory. The exchange correlation functional is the Generalized-Gradient-Approximation (GGA) of the Perdew-Burke-Ernzerhof (PBE) functional. The cutoff energy was set as 450 eV, energy and force convergence criteria set as $10^{-6}$ eV and 0.01 eV Å$^{-1}$. In this work，the relaxed lattice of Mn$_5$Ge$_3$ were a=b=7.122 Å c=4.964 Å that agree with the single crystal experimental lattice constant (a=b=7.19883 Å, c=5.05576 Å). The Ge atoms occupy 6g Wyckoff sites (0.5991, 0, 0.25), the 6g Wyckoff sites (0.2335, 0, 0.25) Mn1 and 4d (1/3, 2/3, 0) Wyckoff sites Mn2 magnetic moment is about 3.1 μB and 2.2 μB. The anomalous Hall conductivity and Berry curvature was Calculated in Wannier90 and WannierTools package. Experimentally, we found that the excess Ge atoms are located randomly in the 2b Wyckoff sites (0, 0, 0). We therefore built the supercell and calculated the band structures of Mn$_5$Ge$_{3+x}$ (x=0.25，0.33，0.5，0.75，and 1) with the extra Ge atoms in the 2d Wyckoff sites (0, 0, 0).

### STEM and L-TEM measurements

The lamellar samples used in STEM and L-TEM observation were prepared by the mechanical polishing procedure. The magnetic field applied normal to the thin plate was induced by the magnetic objective lens of the TEM (JEOL F200). The details of the Lorentz TEM are given in the Supplementary Information.

### Micromagnetic simulations

Micromagnetic simulations are performed by using the Object Oriented MicroMagnetic Framework (OOMMF) code based on the Landau-Lifshitz-Gilbert (LLG) equation. In all simulations, the size of the calculation cell is 1000×1000×100 nm. Magnetic parameters used in the simulation are calculated by using the measurement data: saturation magnetization $M_S$ = 7.7 ×10$^5$ A/m, exchange stiffness A = 3.51×10$^{-11}$ J/m, and perpendicular magnetic anisotropy (PMA) constant $K_u$ = 3.48×10$^5$ J/m$^3$.

## Data availability

The data that support the findings of this study are available from the corresponding author upon reasonable request.

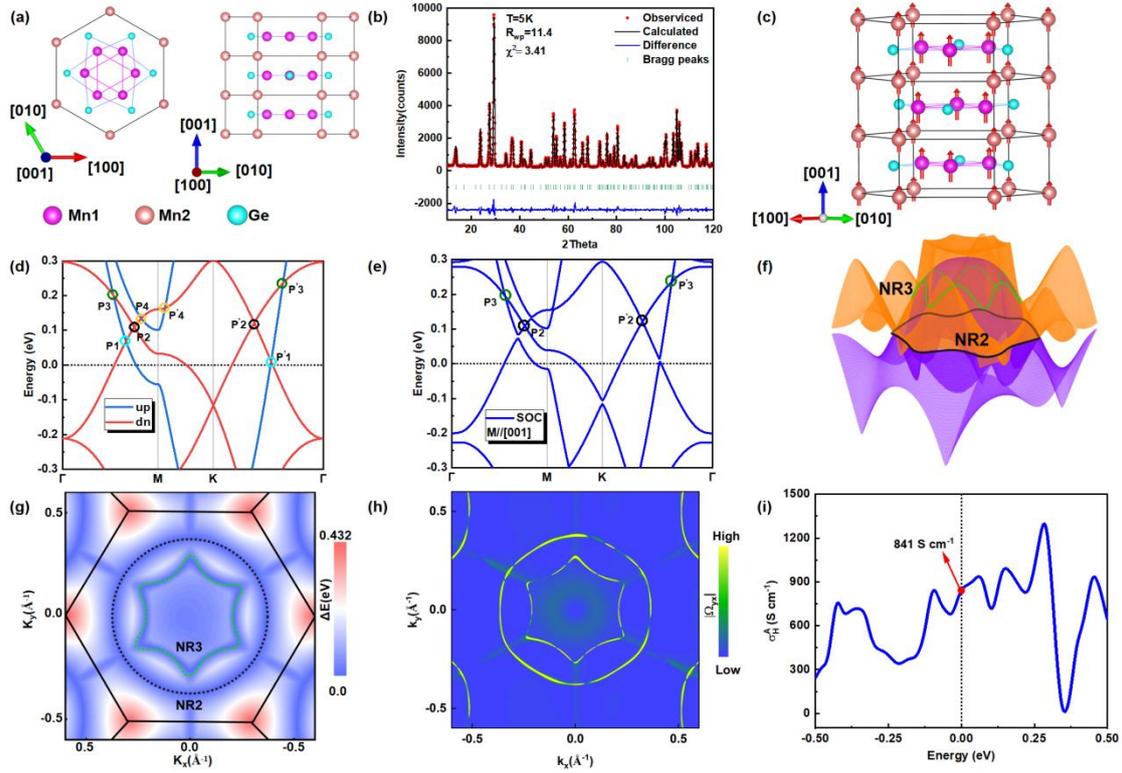

**Fig. 1 Crystal, magnetic and electronic band structures of Mn$_5$Ge$_3$.** (a) Schematic of Mn$_5$Ge$_3$ crystal structure. Purple and orange balls represent Mn1 and Mn2 atoms, green balls represent Ge atoms. (b) Powder neutron diffraction patterns at 5K. Red hollow circles, black and green lines correspond to the observed, calculated and difference patterns, respectively. The green ticks indicate the Bragg peak positions of Mn$_5$Ge$_3$. (c) The refinement magnetic structure of Mn$_5$Ge$_3$ at 5K. (d) The band structure of Mn$_5$Ge$_3$ without considering SOC. The paths of P1-P'1 (cyan circles), P2-P'2 (black circles), P3-P'3 (olive circles) and P4-P'4 (orange circles) represent four Weyl nodal rings. The black dot line represents the Fermi levels. (e) The band structure of Mn$_5$Ge$_3$ considering SOC. Only two Weyl node rings, P2-P'2 (black circles) and P3-P'3 (olive circles), are survived. (f) The 3D and (g) 2D schematic of band structure with Weyl nodal rings. (h) The calculated Berry curvature of the 2D band structure. (i) The calculated AHC of Mn$_5$Ge$_3$ with M parallel to [001] (*c*-axis).

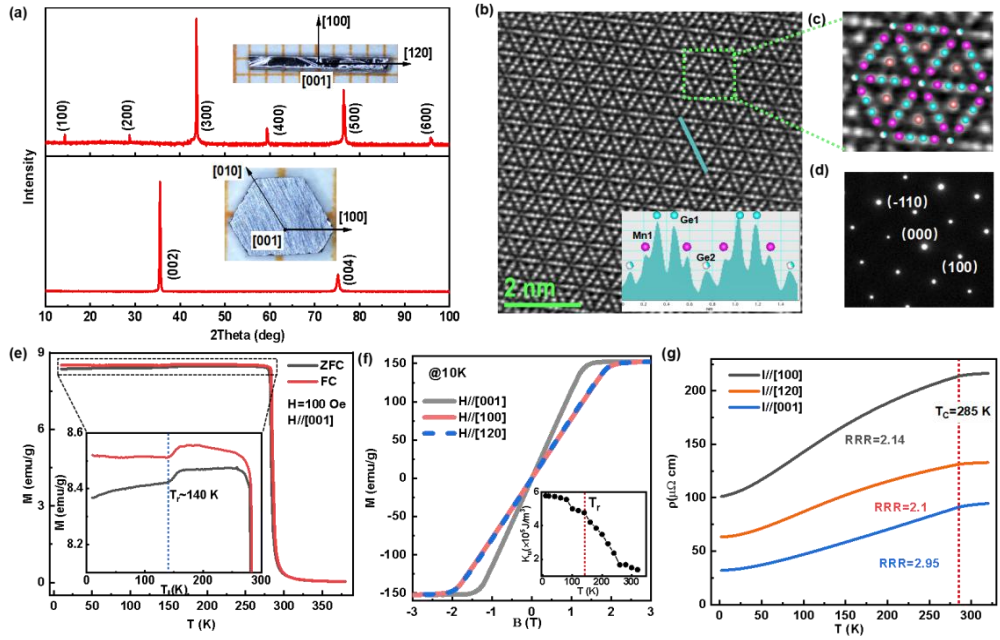

**Fig. 2 Basic crystal, magnetic and electronic transport properties of Mn$_5$Ge$_3$ single crystal.** (a) X-ray diffraction patterns of single-crystal flank (upper) and cross section (bottom). (b) High-resolution STEM image of (001) plane. The inset at the bottom is the profile of the cyan line. The cyan, purple and partially cyan filled balls represent Ge, Mn1 and excess atoms occupying (0,0,0) vacant position. (c) Enlarged view of the green rectangular dashed box in (b). (d) Electron diffraction patterns of (001) plane. (e) Temperature dependence of magnetization curves experienced with ZFC and FC process. (f) Field dependence of magnetization curves along different axis at 10 K. (g) Resistivity curves along different axis at zero filed.

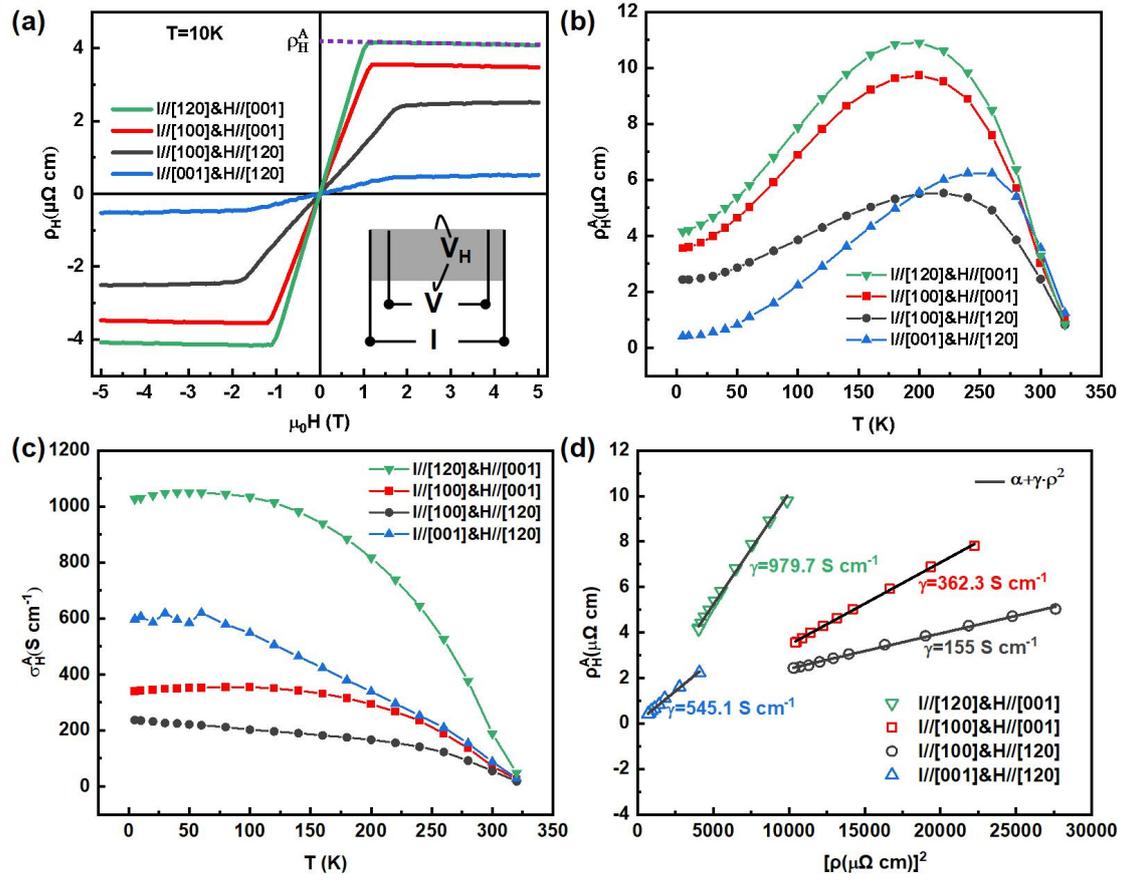

**Fig. 3 Large tunable AHC in $Mn_5Ge_3$ single crystal.** (a) Field dependence of AHC curves along different axes at 10K. The inserts are the schematic diagram of measurements. The purple line represents the linear fitting at saturation region and the intercept is named $\rho_H^A$. (b) Temperature dependent $\rho_H^A$ along different axes. (c) Temperature dependent $\sigma_H^A$ along different axes. (d) The lines for $\rho_H^A$ V.S. $\rho^2$. The black lines are the linear fitting of $\alpha+\gamma\rho^2$.

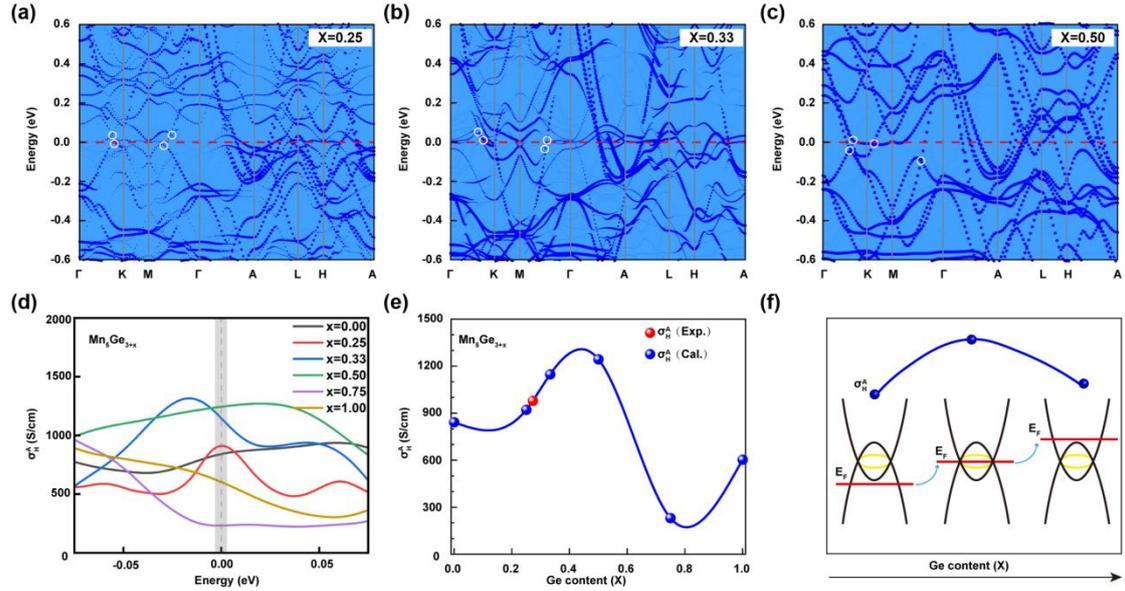

**Fig. 4 The evolution of band structure and corresponding Fermi level of $Mn_5Ge_{3+x}$. (a-c)** bulk band structures of $Mn_5Ge_{3+x}$ for different excess Ge content x=0.25, 0.33 and 0.5, respectively. (d) Energy-dependent AHC of $Mn_5Ge_{3+x}$ for different excess Ge content x. (e) Evolution of the AHC as a function of excess Ge content x from the calculations and experiments (red ball). (f) Schematic plot showing the modulation effects of the band structure and the resultant AHC with different Ge content x. SOC was included for all calculations.

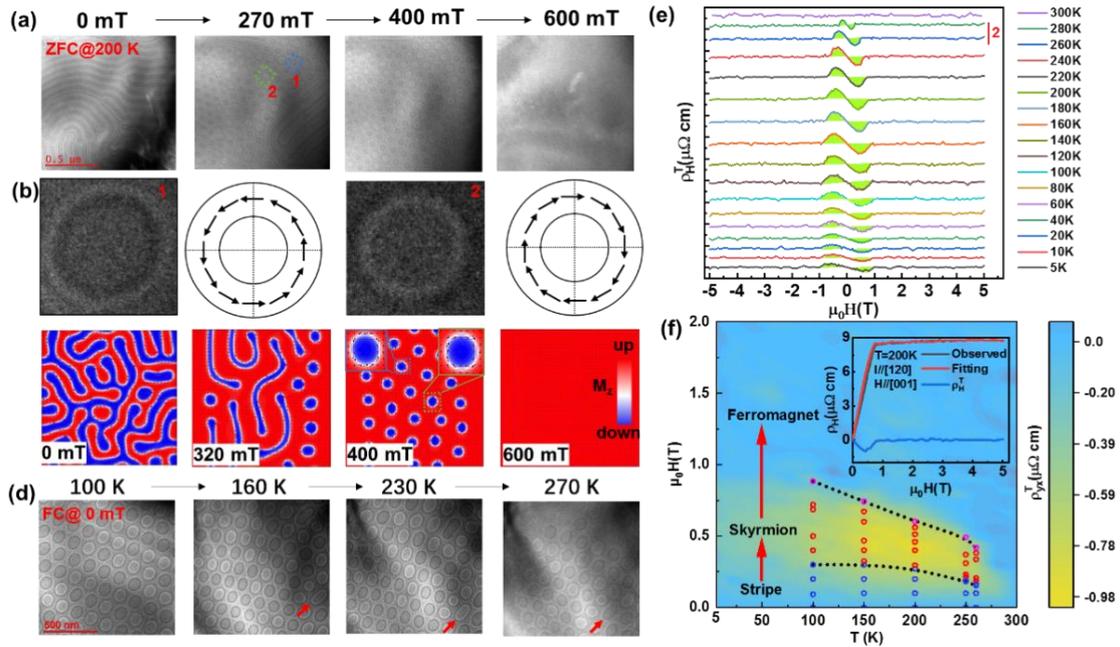

**Fig. 5 Magnetic spin textures and topological Hall effect of Mn$_5$Ge$_3$ single crystal.** (a) The under-focus L-TEM images taken at 100 K with increasing filed along *c* axis. The red dotted box shows the transformation of stipes domains into skyrmion. The red numbers locate at three different bubbles. (b) The under-focus L-TEM and OOMMF simulation images of two opposite chirality skyrmions. (c) The simulation L-TEM images and their schematic of spin arrangement. (d) The under-focus L-TEM images taken at 100, 160, 230, and 270K, respectively, after the FC process with a small field of 500 Oe along [001]. The red arrows show a spontaneous reversal process of skyrmion chirality. (e) The THE curves extracted from the total Hall resistivity curves at various temperatures with I along [120] and H along [001]. (f) The magnetic phase diagram with temperature and fields. The dotted lines and hollow circles represent critical fields from THE curves and L-TEM experiments, respectively. The inset shows the process of extracting THE at 200 K.

# Reference

1. He Q.L., Hughes T.L., Armitage N.P., Tokura Y., Wang K.L. Topological spintronics and magnetoelectronics. *Nat. Mater.* **21**, 15-23 (2022).

2. Bernevig B.A., Felser C., Beidenkopf H. Progress and prospects in magnetic topological materials. *Nature* **603**, 41-51 (2022).

3. Nagaosa N., Sinova J., Onoda S., MacDonald A.H., Ong N.P. Anomalous Hall effect. *Rev. Mod. Phys.* **82**, 1539-1592 (2010).

4. Xiao D., Chang M.-C., Niu Q.. Berry phase effects on electronic properties. *Rev. Mod. Phys.* **82**, 1959-2007 (2010).

5. Nakatsuji S., Kiyohara N., Higo T. Large anomalous Hall effect in a non-collinear antiferromagnet at room temperature. *Nature* **527**, 212-215 (2015).

6. Nayak A.K.*, et al.* Large anomalous Hall effect driven by a nonvanishing Berry curvature in the noncolinear antiferromagnet $Mn_3Ge$. *Sci. Adv.* **2**, e1501870 (2016).

7. Ye L.*, et al.* Massive Dirac fermions in a ferromagnetic kagome metal. *Nature* **555**, 638-642 (2018).

8. Liu E.*, et al.* Giant anomalous Hall effect in a ferromagnetic Kagome-lattice semimetal. *Nat. Phys.* **14**, 1125-1131 (2018).

9. Kim K.*, et al.* Large anomalous Hall current induced by topological nodal lines in a ferromagnetic van der Waals semimetal. *Nat. Mater.* **17**, 794-799 (2018).

10. Manna K.*, et al.* From Colossal to Zero: Controlling the Anomalous Hall Effect in Magnetic Heusler Compounds via Berry Curvature Design. *Phys. Rev. X* **8**, 041045 (2018).

11. Sakai A.*, et al.* Giant anomalous Nernst effect and quantum-critical scaling in a ferromagnetic semimetal. *Nat. Phys.* **14**, 1119-1124 (2018).

12. Belopolski I.*, et al.* Discovery of topological Weyl fermion lines and drumhead surface states in a room temperature magnet. *Science* **365**, 1278-1281 (2019).

13. Li P., *et al.* Giant room temperature anomalous Hall effect and tunable topology in a ferromagnetic topological semimetal $Co_2MnAl$. *Nat. Commun.* **11**, 3476 (2020).

14. Guin S.N., *et al.* 2D-Berry-Curvature-Driven Large Anomalous Hall Effect in Layered Topological Nodal-Line MnAlGe. *Adv. Mater.* **33**, 2006301 (2021).

15. Singh S., *et al.* Anisotropic Nodal-Line-Derived Large Anomalous Hall Conductivity in ZrMnP and HfMnP. *Adv. Mater.* **33**, 2104126 (2021).

16. Mühlbauer S., *et al.* Skyrmion Lattice in a Chiral Magnet. *Science* **323**, 915 (2009).

17. Yu X.Z., *et al.* Near room-temperature formation of a skyrmion crystal in thin-films of the helimagnet FeGe. *Nat. Mater.* **10**, 106-109 (2011).

18. Wang W., *et al.* A Centrosymmetric Hexagonal Magnet with Superstable Biskyrmion Magnetic Nanodomains in a Wide Temperature Range of 100-340 K. *Adv. Mater.* **28**, 6887-6893 (2016).

19. Hou Z., *et al.* Observation of Various and Spontaneous Magnetic Skyrmionic Bubbles at Room Temperature in a Frustrated Kagome Magnet with Uniaxial Magnetic Anisotropy. *Adv. Mater.* **29**, 1701144 (2017).

20. Ding B., *et al.* Observation of Magnetic Skyrmion Bubbles in a van der Waals Ferromagnet $Fe_3GeTe_2$. *Nano Lett* **20**, 868-873 (2020).

21. Maraytta N., *et al.* Anisotropy of the magnetocaloric effect: Example of $Mn_5Ge_3$. *J. Appl. Phys.* **128**, 103903 (2020).

22. Zeng C., Yao Y., Niu Q., Weitering H.H. Linear Magnetization Dependence of the Intrinsic Anomalous Hall Effect. *Phys. Rev. Lett.* **96**, 037204 (2006).

23. M. dos Santos Dias N., *et al.* Topological magnons driven by the Dzyaloshinskii-Moriya interaction in the centrosymmetric ferromagnet $Mn_5Ge_3$. *Nat. Commun.* **14**, 7321 (2023).

24. Reiff W.M., Narasimhan K.S.V.L., Steinfink H. Mössbauer and magnetic investigation of the system $Mn_{5-x}Fe_xGe_3$ (x=0.5, 1.0 and 1.5). *J. Solid State Chem.* **4**, 38-45 (1972).

25. Kappel G., Fischer G., Jaéglé A.. Magnetic investigation of the system Mn5Ge3-Mn5Si3. *phys.status solidi (a)* **34**, 691-696 (1976).

26. Bara J.J., Gajič B.V., Pędziwiatr A.T., Szytuła A. Investigations of crystal and magnetic properties of the $Mn_{5-x}Fe_xGe_3$ intermetallic compounds. *J. Magn. Magn. Mater.* **23**, 149-155 (1981).

27. Gajdzik M., Sürgers C., Kelemen M.T., Löhneysen H.v. Strongly enhanced Curie temperature in carbon-doped $Mn_5Ge_3$ films. *J. Magn. Magn. Mater.* **221**, 248-254 (2000).

28. Songlin, Dagula, Tegus O., Brück E., de Boer F.R., Buschow K.H.J. Magnetic and magnetocaloric properties of $Mn_5Ge_{3-x}Sb_x$. *J. Alloys Compd.* **337**, 269-271 (2002).

29. Panguluri R.P., Zeng C., Weitering H.H., Sullivan J.M., Erwin S.C., Nadgorny B. Spin polarization and electronic structure of ferromagnetic $Mn_5Ge_3$ epilayers. *phys. status solidi (b)* **242**, R67-R69 (2005).

30. Zeng C., *et al.* Epitaxial ferromagnetic $Mn_5Ge_3$ on Ge(111). *Appl. Phys. Lett.* **83**, 5002-5004 (2003).

31. Tang J., *et al.* Electrical Probing of Magnetic Phase Transition and Domain Wall Motion in Single-Crystalline $Mn_5Ge_3$ Nanowire. *Nano Lett* **12**, 6372-6379 (2012).

32. Tang J., *et al.* Electrical Spin Injection and Detection in $Mn_5Ge_3$/Ge/$Mn_5Ge_3$ Nanowire Transistors. *Nano Lett* **13**, 4036-4043 (2013).

33. Castelliz L. Kristallstruktur von $Mn_5Ge_3$ und einiger ternärer Phasen mit zwei Übergangselementen. *Monatshefte für Chemie und verwandte Teile anderer Wissenschaften* **84**, 765-776 (1953).

34. Ohoyama T. X-ray and Magnetic Studies of the Manganese-Germanium System. *J. Phys. Soc. Jpn.* **16**, 1995-2002 (1961).

35. Tawara Y., Sato K. On the Magnetic Anisotropy of Single Crystal of $Mn_5Ge_3$. *J. Phys. Soc. Jpn.* **18**, 773-777 (1963).

36. Aharoni A. Demagnetizing factors for rectangular ferromagnetic prisms. *J. Appl. Phys.* **83**, 3432-3434 (1998).

37. Ciszewski R. Magnetic Structure of the $Mn_5Ge_3$ Alloy. *phys. status solidi (b)* **3**, 1999-2004 (1963).

38. Tian Y., Ye L., Jin X.. Proper scaling of the anomalous Hall effect. *Phys. Rev. Lett.* **103**, 087206 (2009).

39. Shen J*., et al.* Local Disorder-Induced Elevation of Intrinsic Anomalous Hall Conductance in an Electron-Doped Magnetic Weyl Semimetal. *Phys. Rev. Lett.* **125**,  (2020).

40. Ding B*., et al.* Manipulating Spin Chirality of Magnetic Skyrmion Bubbles by In-Plane Reversed Magnetic Fields in $(Mn_{1-x}Ni_x)_{65}Ga_{35}$ (x=0.45) Magnet. *Phys. Rev. Appl.* **12**, 054060 (2019).

41. Nagao M*., et al.* Direct observation and dynamics of spontaneous skyrmion-like magnetic domains in a ferromagnet. *Nat. Nanotechnol.* **8**, 325-328 (2013).

42. Singh A*., et al.* Scaling of domain cascades in stripe and skyrmion phases. *Nat. Commun.* **10**, 1988 (2019).

43. Wang X.R., Hu X.-C., Sun Z.-Z. Topological Equivalence of Stripy States and Skyrmion Crystals. *Nano Lett* **23**, 3954-3962 (2023).

## Acknowledgements

This work was supported by the National Key R&D program of China (No. 2022YFA1402600), National Natural Science Foundation of China (Grants Nos. 12204347, 12274321, 12074415 and 12361141823) and Beijing National Laboratory for Condensed Matter Physics (2023BNLCMPKF011). A portion of this work was carried out at the Synergetic Extreme Condition User Facility (SECUF) in Huairou Science City.

## Author contributions

W.H.W. conceived the idea and supervised the overall research. H.L. carried out the preparation of the $Mn_5Ge_3$ single crystals. B.D., Y.Y., and H.L. performed the LTEM and STEM experiments. F. Z. and H. L. performed the ab initio calculation. W.Y.Y, and J.B.Y performed the powder neutron experiments. J.C., L.X.S., X.K.X.,Y.-C.L.,Y.

L., and Y. J. participated in the discussion of the article. The manuscript was drafted by W.H.W., H.L., F.Z., and B.D. All authors discussed the results and contributed to the manuscript preparation.

## Competing interests

The authors declare no competing interests.

## Additional information

**Supplementary information** The online version contains supplementary material available at online.

**Correspondence** and requests for materials should be addressed to Wenhong Wang

# Supplementary information for

# Coexistence of large anomalous Hall effect and topological magnetic skyrmions in a Weyl nodal ring ferromagnet $Mn_5Ge_3$

Hang Li[1,5], Feng Zhou[1,5], Bei Ding[2,5], Jie Chen[1], Linxuan Song[2], Wenyun Yang[4], Yong-Chang Lau[2], Jinbo Yang[4], Yue Li[1], Yong Jiang[1], Wenhong Wang[1,2]*

[1]Institute of Quantum Materials and Devices, School of Electronics and Information Engineering, Tiangong University, Tianjin 300387, China
[2]Institute of Physics, Chinese Academy of Sciences, Beijing 100190, China
[3]University of Chinese Academy of Sciences, Beijing 100049, China
[4]State Key Laboratory for Artificial Microstructure & Mesoscopic Physics, School of Physics, Peking University, Beijing 100871, P. R. China
[5] These authors contributed equally: Hang Li, Feng Zhou, and Bei Ding.
*E-mail: wenhongwang@tiangong.edu.cn

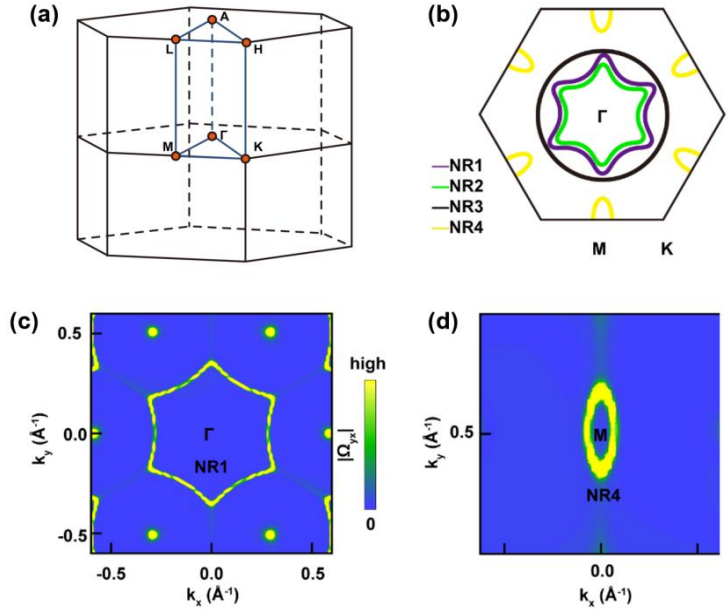

Fig. S1. Calculated nodal rings of Mn$_5$Ge$_3$. (a) $k_z$=0 plan of the band crossing points without SOC. (b) and (c) The calculated Berry curvature of NR1 and NR4 with SOC.

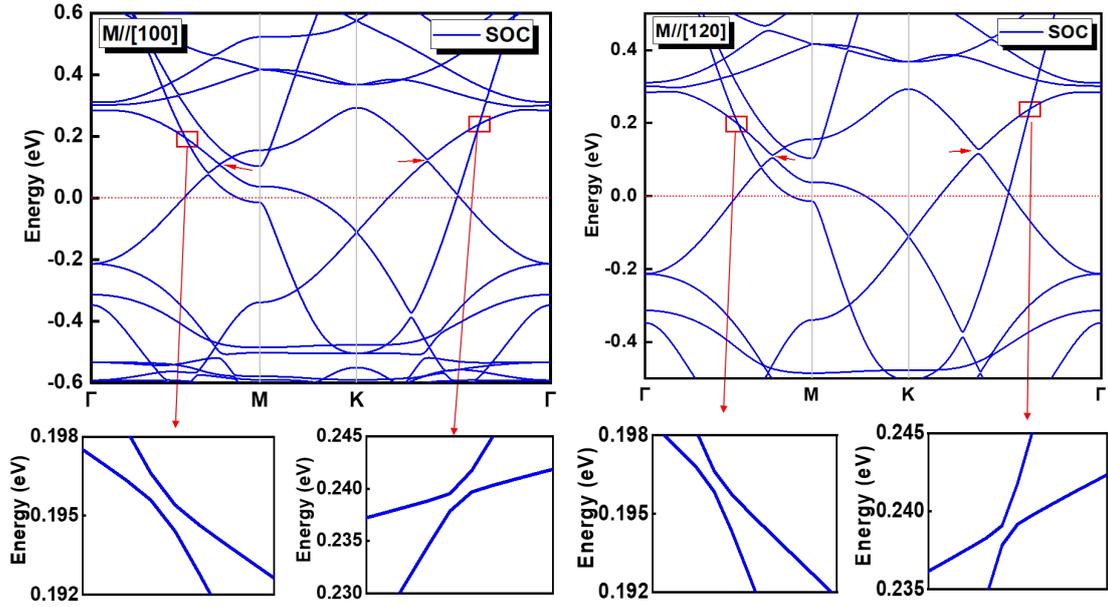

Fig. S2. Calculated band structure of of Mn$_5$Ge$_3$ with the magnetic moments are along the [100] and [120] (*ab*-plane) directions, respectively.

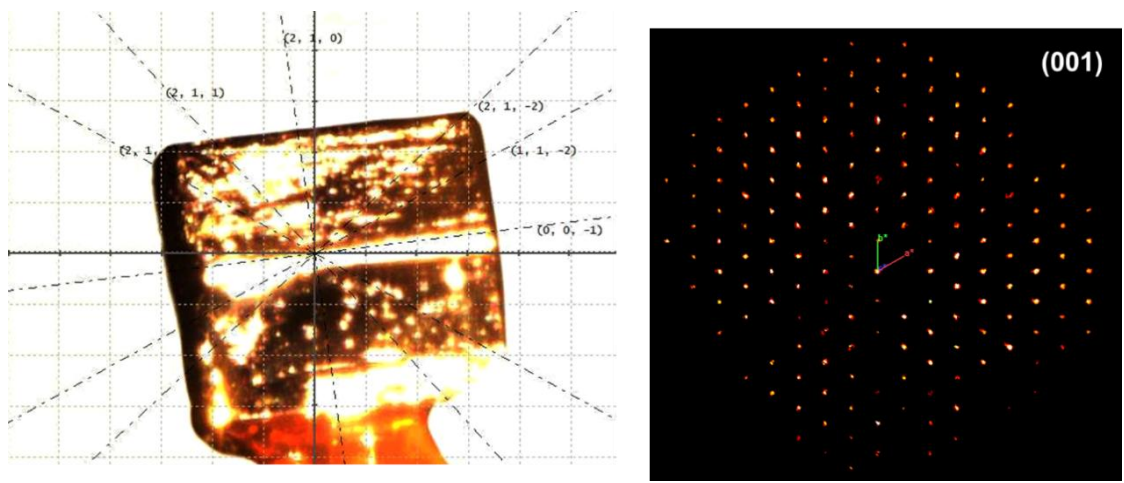

Fig.S3. Mn$_5$Ge$_3$ single-crystal X-ray diffraction pattern.

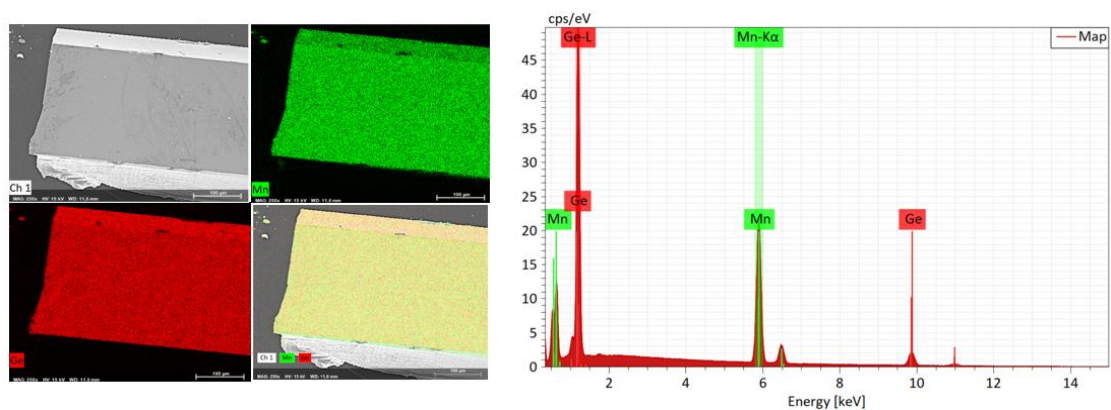

| Sample Number | Mn | Ge | Sample Number | Mn | Ge |
|---|---|---|---|---|---|
| 1# | 60.24 | 39.76 | 4# | 60.49 | 39.51 |
| 2# | 60.07 | 39.93 | 5# | 59.89 | 40.11 |
| 3# | 60.72 | 39.28 | 6# | 60.57 | 39.43 |

Fig. S4. The EDS results of prepared Mn$_5$Ge$_3$ single crystal.

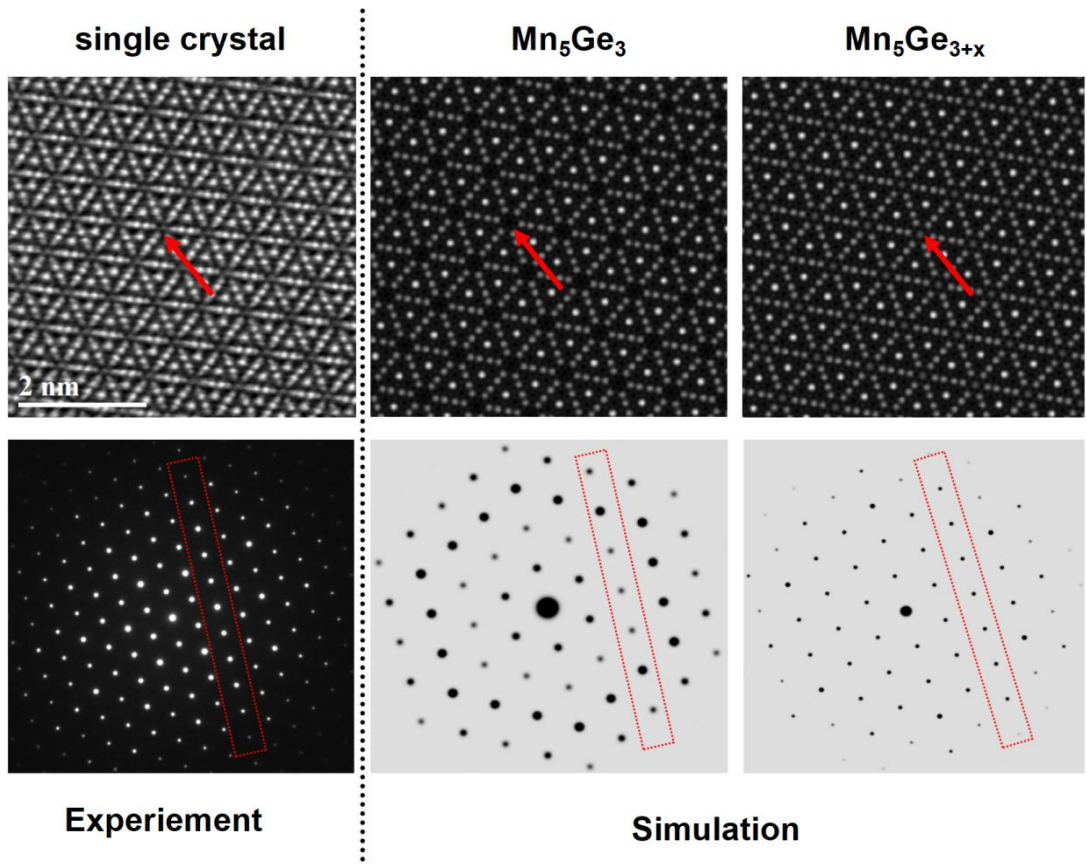

Fig. S5. The atomic STEM and simulation images of *ab* plane.

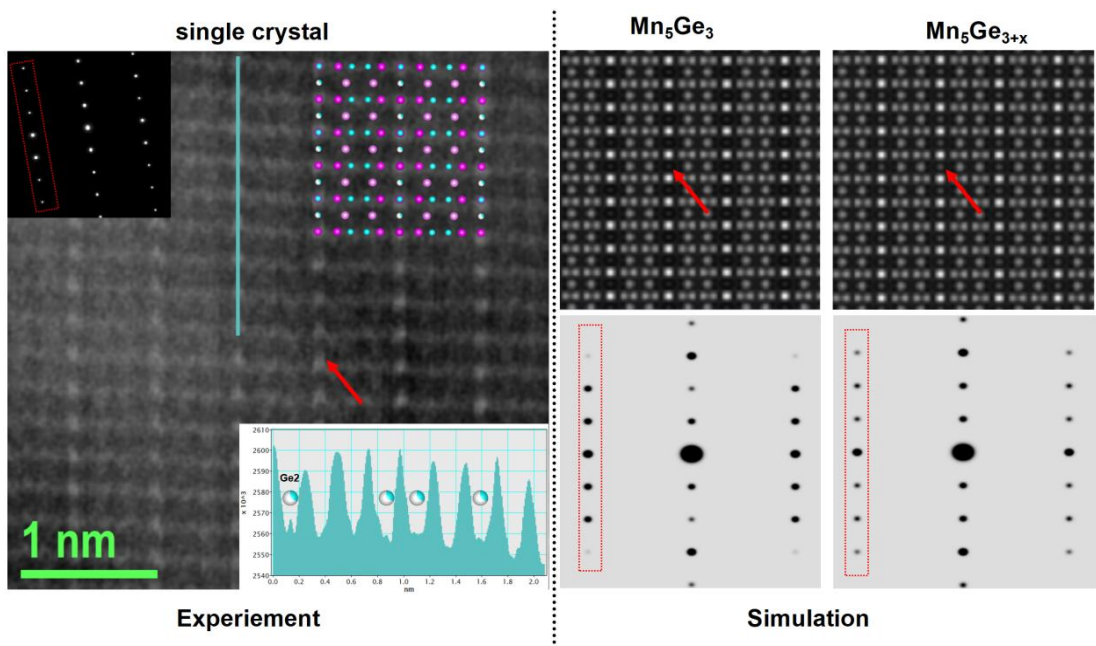

Fig. S6. The atomic STEM and simulation images of *ac* plane.

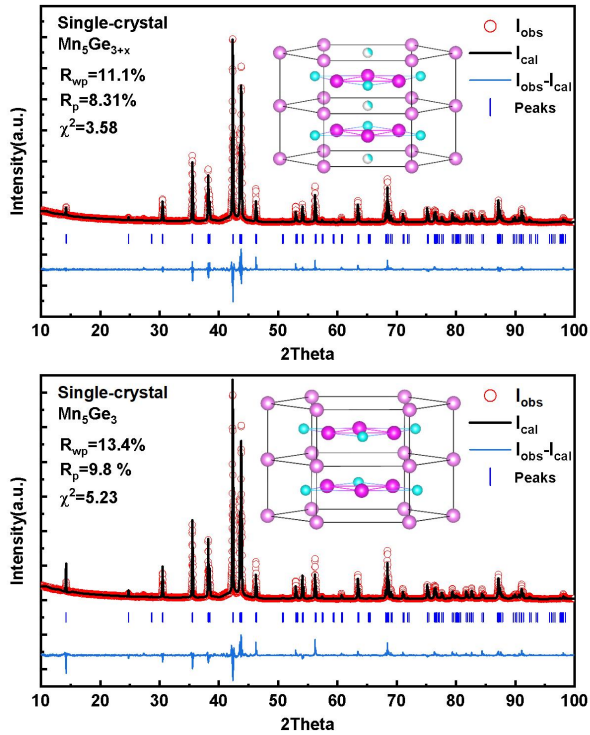

Fig. S7. The Rietveld refinement of the single crystal powder XRD data at room temperature with $Mn_5Ge_{3+x}$ and $Mn_5Ge_3$, respectively.

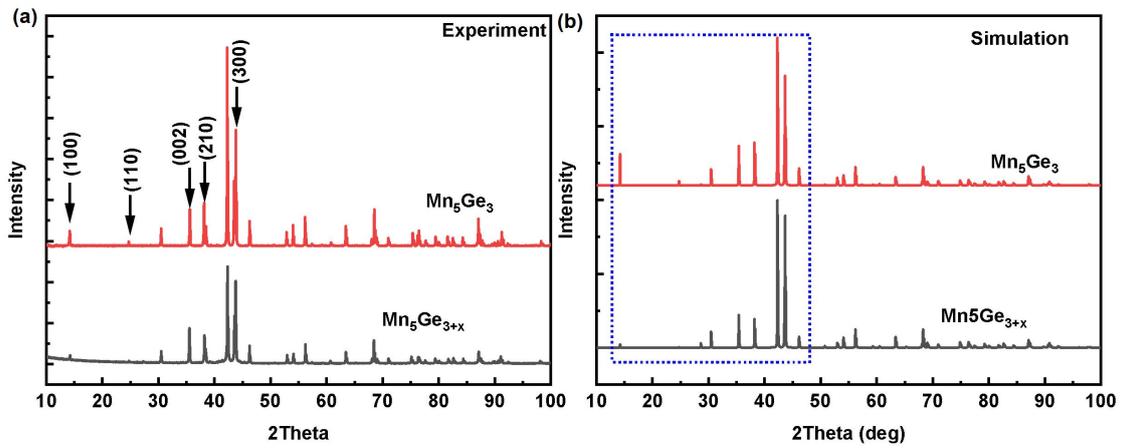

Fig. S8. The experiment and simulation XRD patterns of $Mn_5Ge_{3+x}$ and $Mn_5Ge_3$.

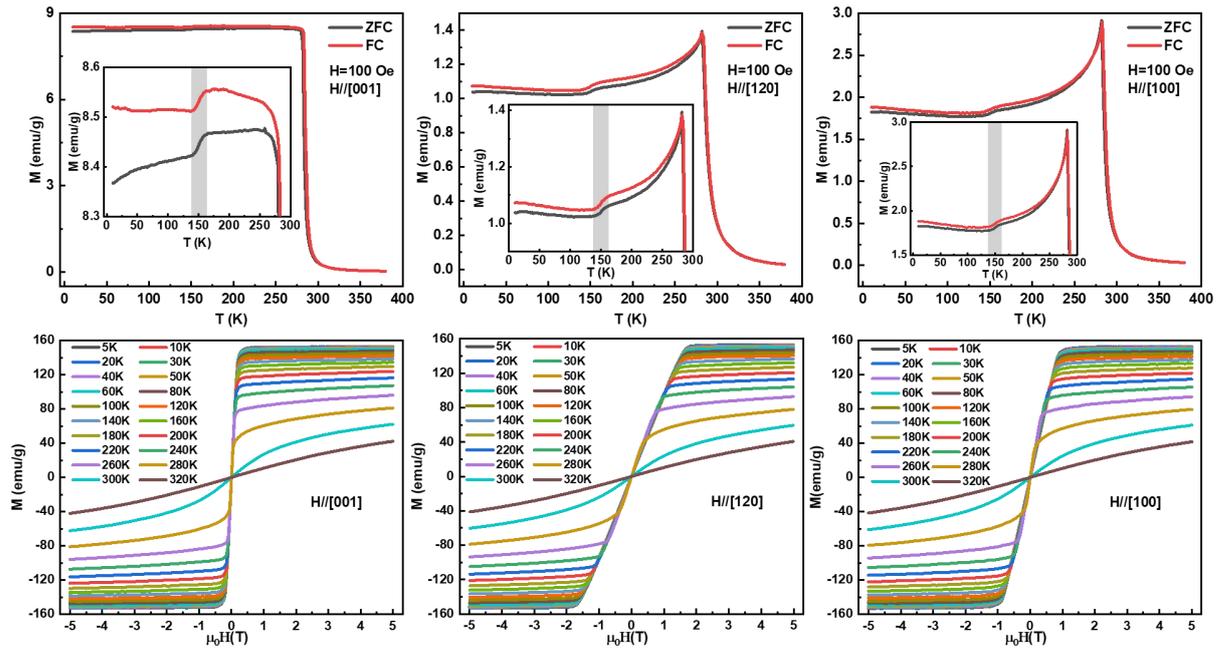

Fig. S9. The M(T) and M(H) curves of Ge-rich Mn$_5$Ge$_3$ single crystal at different axes.

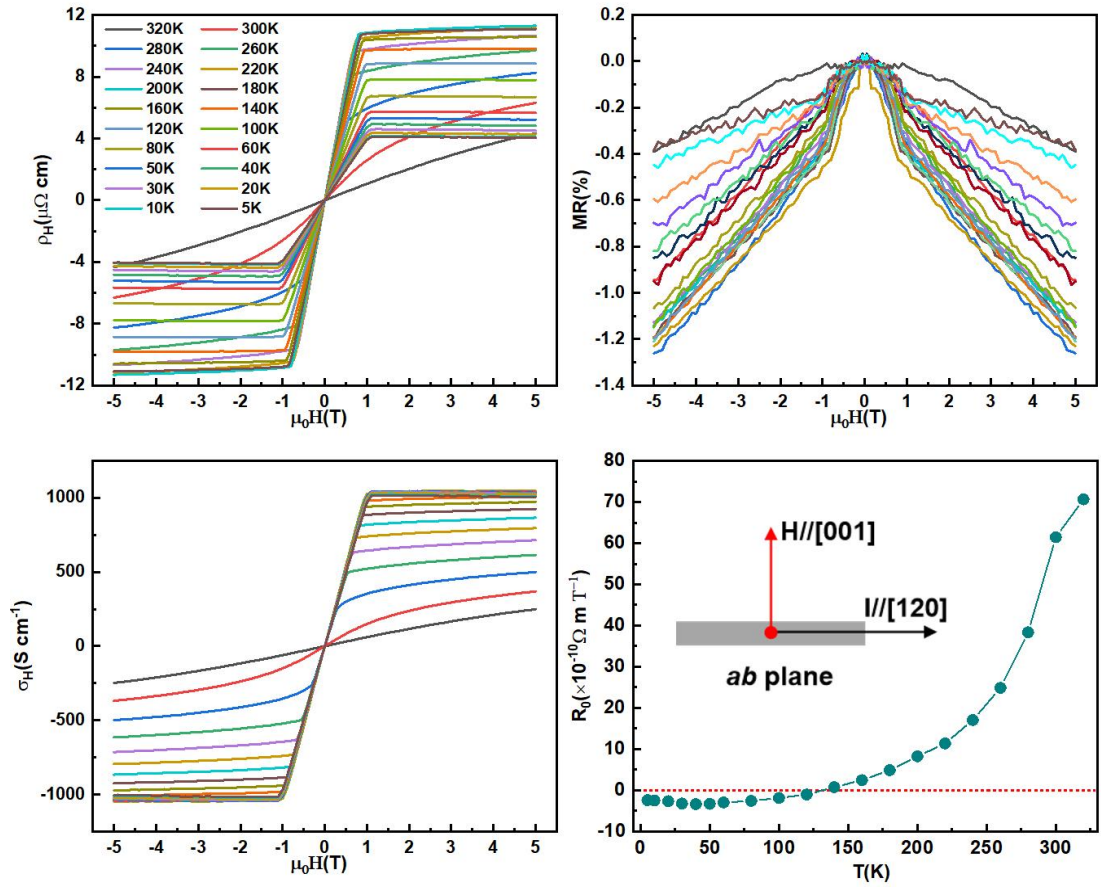

Fig. S10. The detailed anisotropic Hall and Magnetoresistance curves of Ge-rich Mn$_5$Ge$_3$ single crystal with H along [001] and I along [120].

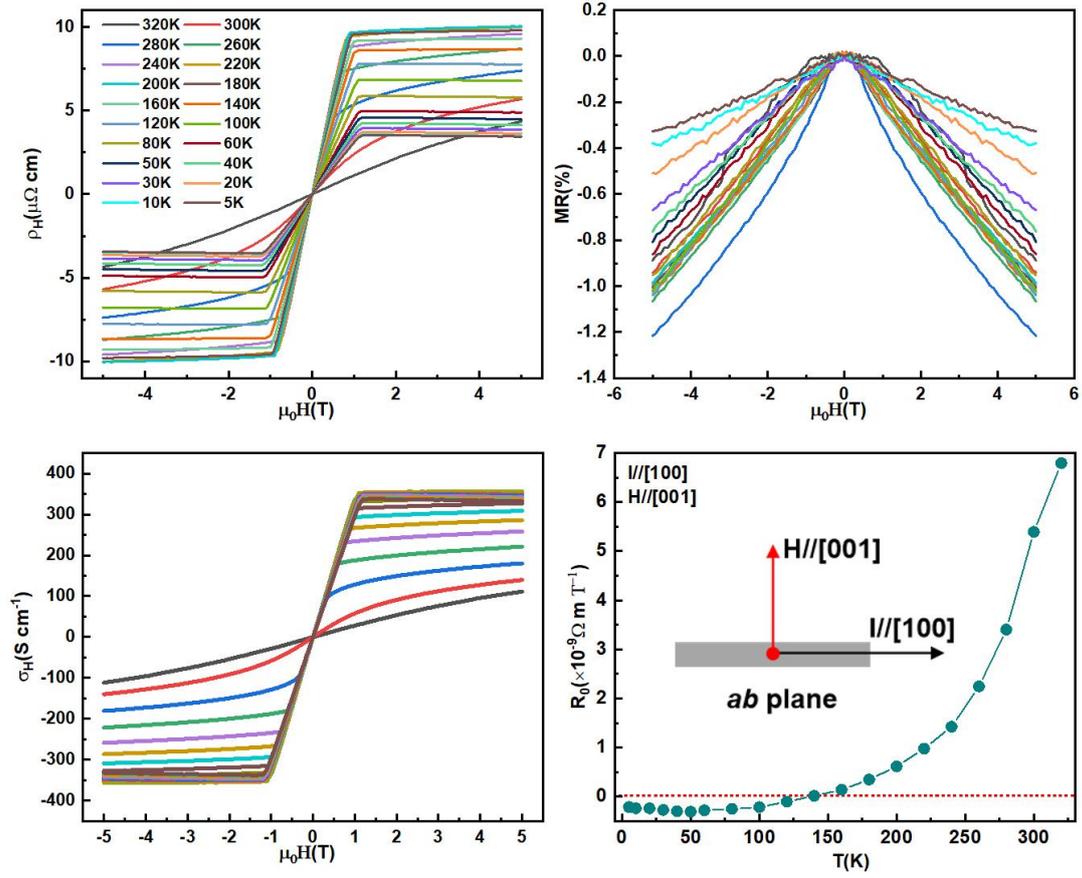

Fig. S11. The detailed anisotropic Hall and Magnetoresistance curves of Ge-rich Mn$_5$Ge$_3$ single crystal with H along [001] and I along [100].

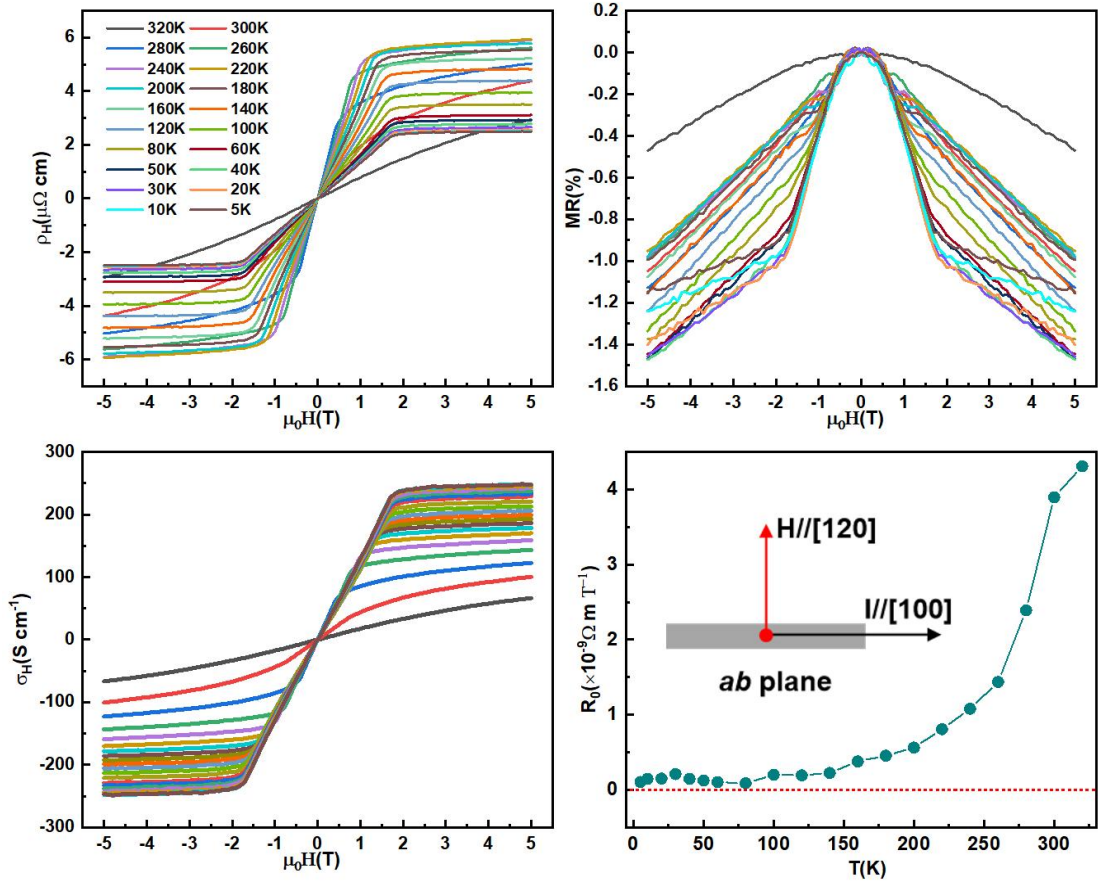

Fig. S12. The detailed anisotropic Hall and Magnetoresistance curves of Ge-rich Mn$_5$Ge$_3$ single crystal with H along [120] and I along [100].

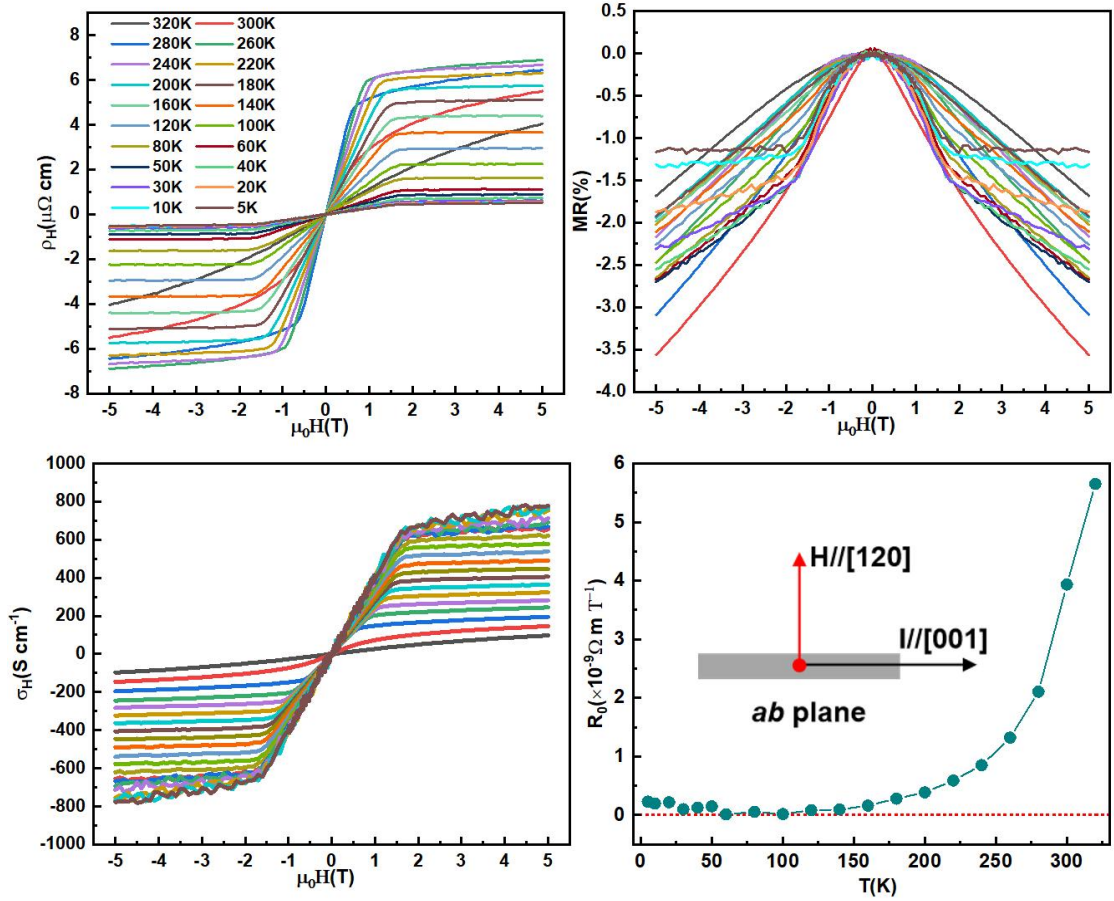

Fig. S13. The detailed anisotropic Hall and Magnetoresistance curves of Ge-rich Mn$_5$Ge$_3$ single crystal with H along [120] and I along [001].

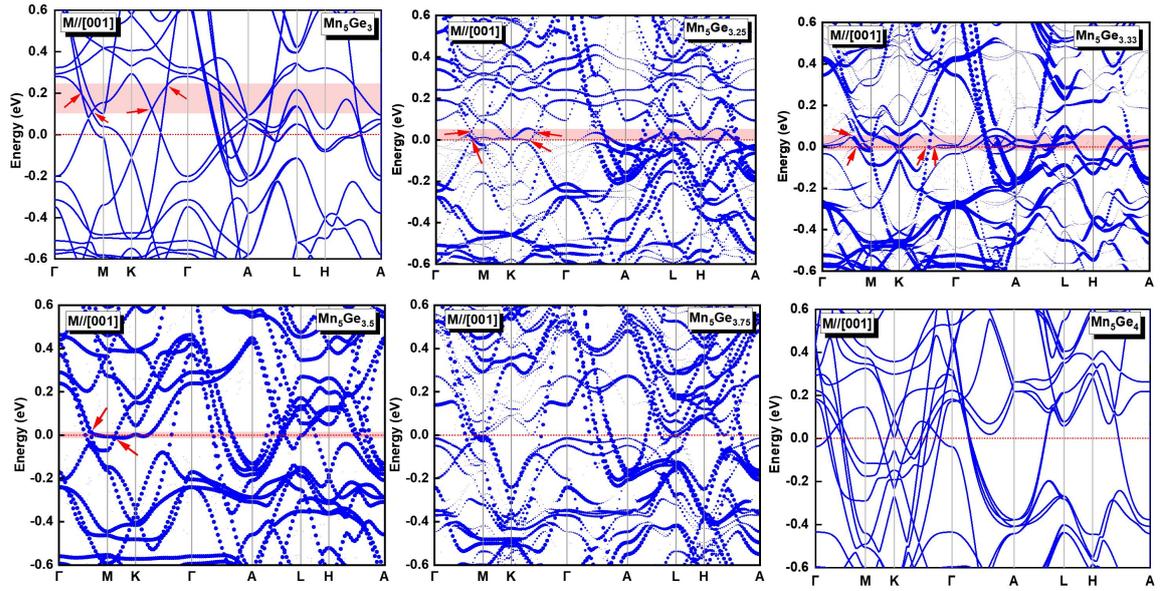

Fig. S14. The calculated band structure of $Mn_5Ge_{3+x}$ with considering SOC along [001]. The red arrows point to the location of the Weyl nodal rings.

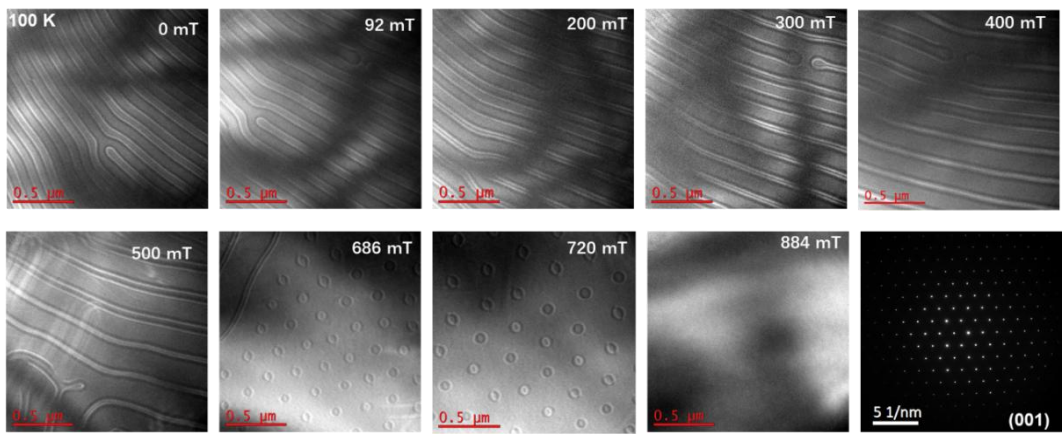
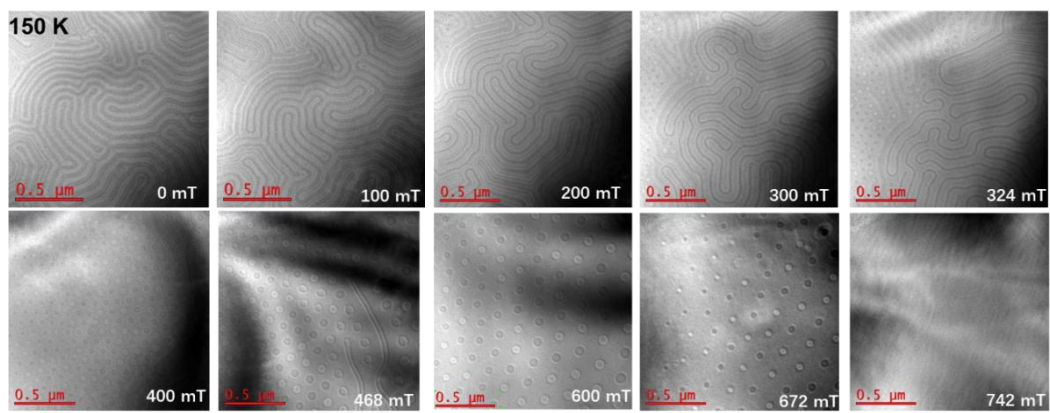
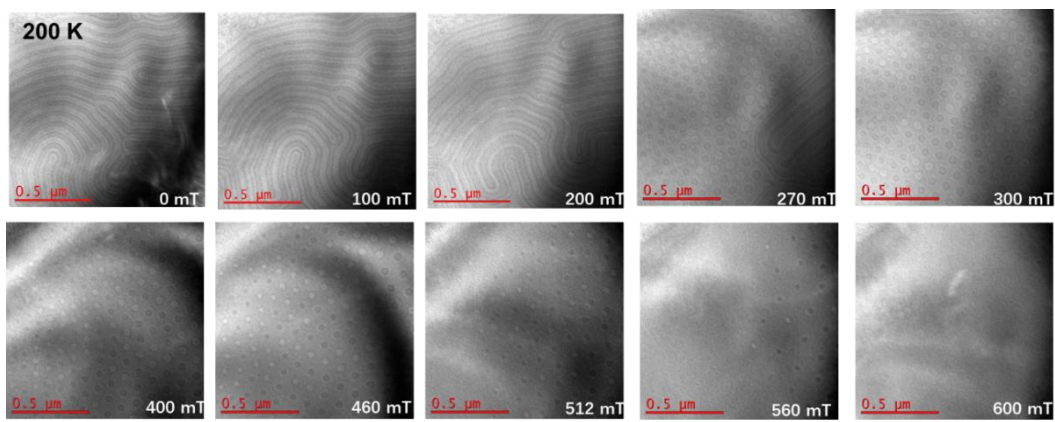
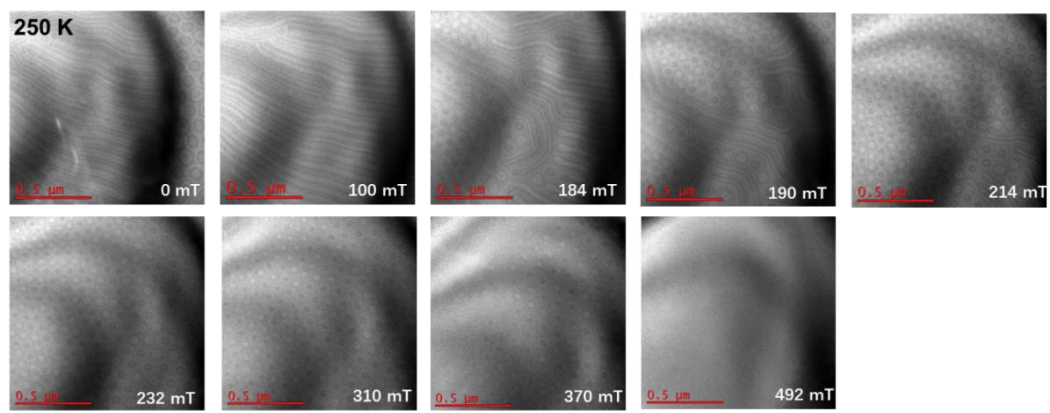

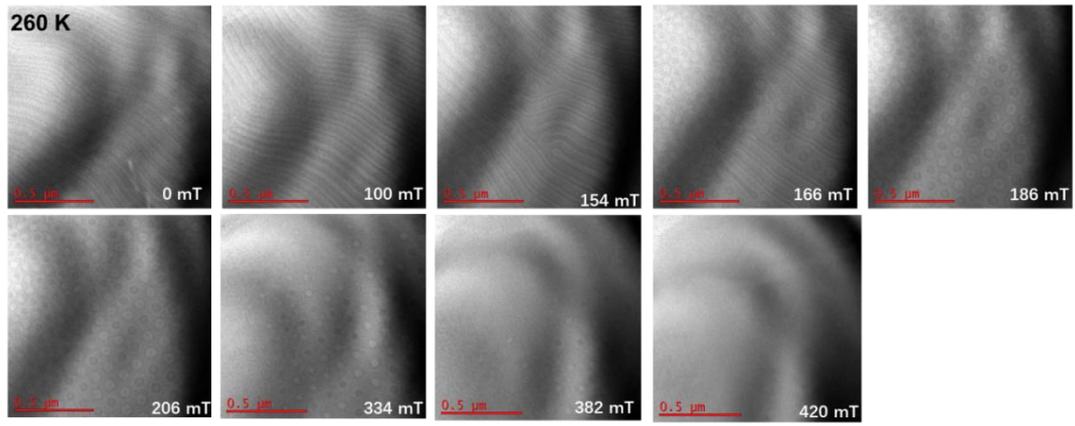

Fig. S15. The detailed LTEM images of Ge-rich $Mn_5Ge_3$ single crystal after ZFC process at different temperature and filed.

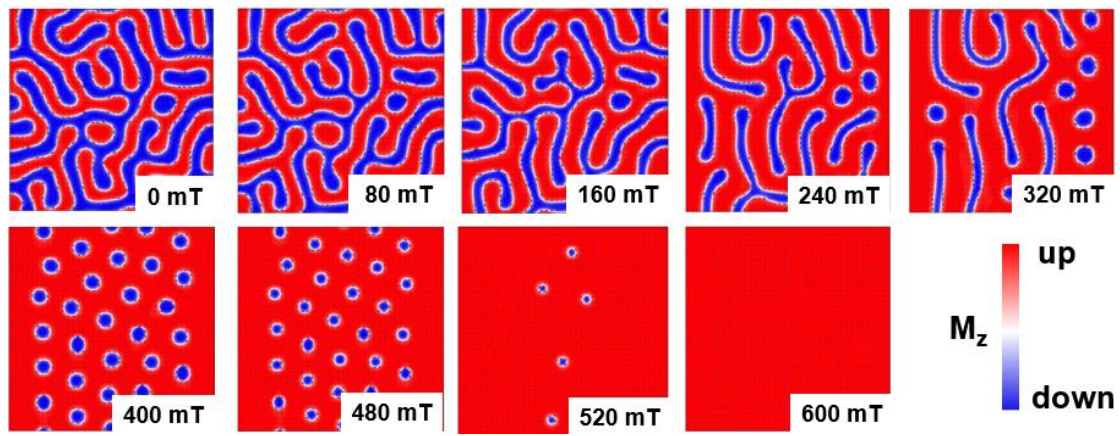

Fig. S16. The OOMMF simulation images using the 200 K experiment data.

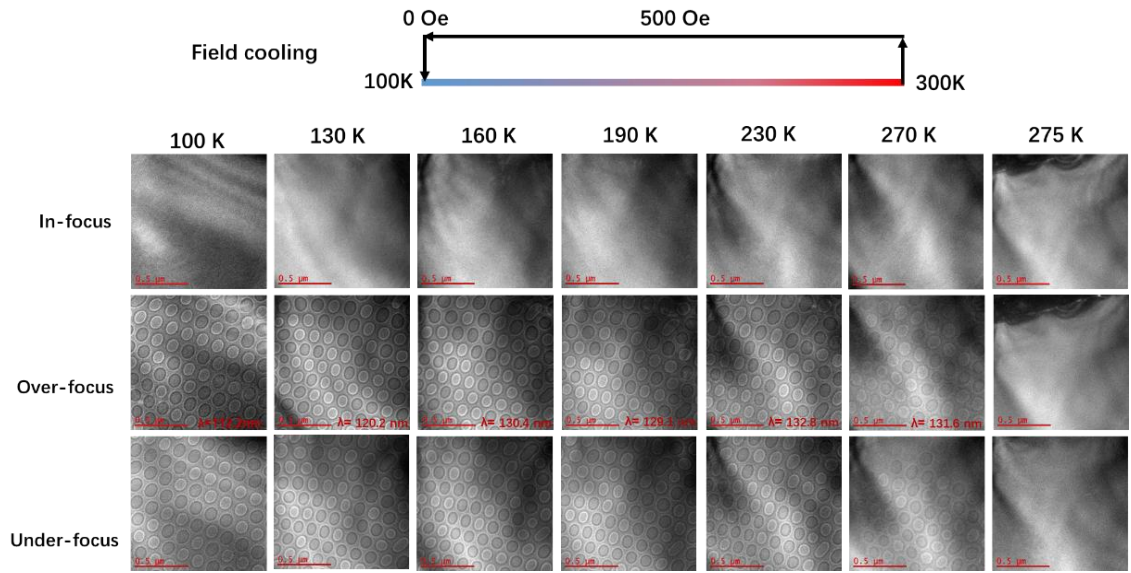

Fig. S17. The LTEM images of Ge-rich $Mn_5Ge_3$ single crystal after 500 Oe FC process.